# Electrostatic deposition of graphene in a gaseous environment: A deterministic route to synthesize rolled graphenes?

*Anton Sidorov, David Mudd, Gamini Sumanasekera, P. J. Ouseph, C. S. Jayanthi, and Shi-Yu Wu\**

Department of Physics, University of Louisville, Louisville, KY 40292, USA

\*Corresponding Author sywu0001@gwise.louisville.edu

The synthesis of single-wall carbon nanotubes (SWCNTs) of desired diameters and chiralities is critical to the design of nanoscale electronic devices with desired properties.[1-6] The existing methods are based on self-assembly,[7-16] therefore lacking the control over their diameters and chiralities. The present work reports a direct route to roll graphene. Specifically, we found that the electrostatic deposition of graphene yielded: (i) flat graphene layers under high vacuum ($10^{-7}$ Torr), (ii) completely scrolled graphene under hydrogen atmosphere, (iii) partially scrolled graphene under nitrogen atmosphere, and (iv) no scrolling for helium atmospheres. Our study shows that the application of the electrostatic field facilitates the rolling of graphene sheets exposed to appropriate gases and allows the rolling of any size graphene. The technique proposed here, in conjunction with a technique that produces graphene nanoribbons (GNRs) of uniform widths, will have significant impact on the development of carbon nanotube based devices. Furthermore, the present technique may be applied to obtain tubes/scrolls of other layered materials.

 

The advancement in carbon nanotube based technology is currently impeded by the lack of control of synthesizing single-wall carbon nanotubes with pre-defined diameters and chiralities and the inherent presence of remnant catalytic particles[1-6]. The existing synthesis techniques, [7-14] including the chemical vapor deposition (CVD)[8], the laser ablation[9], and the arc-discharge[10-12] methods, for producing SWCNTs are stochastic in nature and, therefore, they lead to a distribution of diameters and chiralities for nanotubes. Furthermore, the removal of remnant catalytic nano-particles by methods such as sonication, acid refluxing *etc*.[15,16] may cause inevitable damages to the SWCNTs.

To achieve better control over the diameter and chirality of nanotubes and to circumvent issues associated with remnant catalytic particles, a deterministic route for synthesizing single-wall carbon nanotubes must be sought. Because of the recent progress in making monolayer graphene[17,18] and graphene nano-ribbons (GNRs)[19], the time is ripe to conceive a direct synthesis route for nanotubes. In fact, a recent molecular dynamics study has theoretically demonstrated the feasibility of adsorbate-assisted rolling of patterned graphene nanoribbons on graphite[20].

The premise of this molecular dynamics study is that thin films can be bent by the stress due to the surface adsorption of atomic species. This concept was tested for hydrogen atoms adsorbed on graphene ribbons patterned over a graphite film. More specifically, a 50% random coverage of hydrogen (H) atoms on a GNR of 1.7 nm width patterned on a graphite film was studied, using first principles molecular dynamics (FPMD).[20] It was found that the adsorption of H atoms induces a stress on the GNR, driving it to fold downward and detach from the graphite substrate. Eventually the two edges of the GNR meet and bond together to form a SWCNT. The possibility of synthesizing SWCNTs of various diameters and chiralities were also explored in this study. An upper bound on the radius of the SWCNT that could be synthesized by H adsorption on a single sheet of GNR was estimated to be ~0.7 nm. To synthesize SWCNTs with diameters larger than 0.7 nm, a classical molecular dynamics simulation that considered two sheets of GNRs was proposed.

If the method works as prescribed, it will provide a direct means to synthesize SWCNTs of predefined diameters and chiralities, and remove the concern over catalytic impurities, thus overcoming the major stumbling blocks for realizing the potential for many innovative applications based on SWCNTs. The theoretically designed procedure involves two key steps, namely, the lifting of a van der Waals force-bonded GNR from the substrate and subsequently the bending of the two edges of the GNR so that they meet to form a SWCNT. The key principle behind the steps considered in the theoretical MD simulations is the stress induced by the adsorption of appropriate atoms on a GNR. It is therefore of utmost importance to experimentally validate this scenario. In addition, the suggestion of using two patterned GNRs to synthesize SWCNTs of radius > 0.7 nm may not be realistically feasible. Hence other means to enhance the lifting of a GNR from the substrate needs to be examined.

An experimental route for direct rolling of graphene must involve a well-conceived procedure for obtaining monolayer graphene. Techniques for obtaining monolayer graphene, few-layer graphene, multi-layer graphene, uniform and non-uniform width GNRs have been developed recently.[19, 21-26] These include mechanical exfoliation of highly oriented pyrolitic graphite (HOPG) using either micromachining or the scotch tape technique,[21,22] epitaxial growth of graphene layers on a substrate[23], chemical exfoliations of intercalated graphites[24] or graphite oxide sheets[25], and the electrostatic deposition technique for transferring graphene layers to a substrate[26]. Furthermore, multi-layer graphene nanoscrolls have also been produced by high energy ball milling of graphite powder[27] and by exfoliation of potassium intercalated graphite[28].

In this Letter, we report our observations on rolled/scrolled graphene sheets formed during the electrostatic deposition of graphene under selected gaseous environments. We employ our recently developed technique for electrostatic deposition of graphene layers on a substrate[26], where the electrostatic field assists in lifting the loosely bonded graphene layers found on a freshly cleaved highly pyrolitic graphite (HOPG). By adjusting the strength of the electrostatic field, monolayer to few-layer depositions of graphene on a substrate have been achieved. In the present work we demonstrate that, if the electrostatic deposition process is carried out under selected gaseous atmospheres, depending on the gaseous environment, either complete rolling, partial rolling, or no rolling of graphene can be obtained. In particular, it will be shown that the combined effects due to forces associated with the applied electrostatic field and the stress due to adsorbate atoms play a pivotal role in the rolling/scrolling of any size GNR.

The gaseous environments considered in our study include: (i) hydrogen, (ii) helium, and (iii) nitrogen. For characterizations of rolled/scrolled graphenes, we used both atomic force microscopy (AFM) and high-resolution transmission electron microscopy (HRTEM) in our investigation.

The experimental setup used for the electrostatic deposition is shown in Fig. 1. In the setup, one side of a HOPG sample is glued to a copper electrode with silver epoxy. The other side of the HOPG sample is cleaved using the scotch tape technique to obtain a clean surface. The second electrode is a 3-mm thick copper plate with a 0.1 mm thick mica sheet placed on it. The substrate of interest is then placed over the mica sheet. The electrodes were connected to the terminals of a high voltage source (0-30 kV and 0-10 mA, ER series, Glassman High Voltage, Inc.) via high voltage vacuum feedthroughs. This setup was placed in a chamber consisting of a 4-way stainless cross connected to an ion pump and a gas handling system. Before introducing the gas, the chamber was pumped to a vacuum better than $10^{-7}$ Torr using the ion pump. Then, the pump was isolated and the desired gas (hydrogen, helium, or nitrogen) was introduced. The electrostatic deposition was performed under a pressure of 50 Torr. The maximum applied voltage that can be applied was limited by the breakdown voltage corresponding to each gas.

In our study, two kinds of substrates were used: (i) 300 nm thick $SiO_2$ layer on a silicon substrate (500 $\mu$m thick) suitable for structural characterizations using AFM and (ii) copper TEM grids (coated with holey carbon) suitable for structural characterizations using HRTEM.

To elucidate the interplay between the applied electrostatic field and the gaseous environment on rolling/scrolling of graphene, we have studied: (i) graphene layer deposited in high vaccum, (ii) graphene layer subject to the stress arising from gaseous environments only, (iii) graphene layer subject to the electrostatic field only, and (iv) graphene layer subject to both electrostatic and gaseous environments.

In Fig. 2(a), the atomic force microscope (AFM) image of a monolayer graphene sheet deposited under high vacuum (~$10^{-7}$ torr) is shown. In Fig. 2(b), the AFM image of the same monolayer sheet exposed to hydrogen at ~50 torr is shown. It can be seen that both images are very flat, with no indication that the image corresponding to the one exposed to hydrogen exhibits any effect of rollup. Since the size of the monolayer graphene sheet (~ 12 μm X 22 μm ) is considerably greater than the size corresponding to the theoretical upper limit of the width of a GNR for the formation of a SWCNT, it is not surprising that the adsorption of hydrogen atoms on the graphene sheet is not sufficient to lift the sheet from the substrate. This observation of course does not necessarily suggest any contradiction with the results of the MD simulations[20]. It simply means that the stress effects from adsorbed gases are not sufficient to roll the graphene sheet under consideration and that additional forces may be required to lift and role that sheet. Since loosely bonded graphene sheets on a HOPG sample can be easily deposited to a substrate by an electrostatic field, the application of an electrostatic field may be sufficient to enhance the force on the hydrogen-exposed graphene sheet to dislodge it from the substrate. To test this idea, we studied the deposition of a monolayer graphene sheet exposed to the hydrogen gas to a substrate using an electrostatic field, and compared this study to the case where only an electrostatic field is applied. In addition, we have also studied and compared the cases where the electrostatic depositions were carried out in different gaseous environments. Such a comparison can clearly delineate the roles played by different factors. In Fig. 3, the AFM image of the hydrogen-exposed graphene sheet under an applied voltage of 450 V is shown. It can be seen from Fig. 3 that the hydrogen-exposed monolayer graphene sheet under the application of an electrostatic field is rolled up. The inset in each figure shows the schematic of the scrolled graphene layer for clarity. The degree of scrolling in each case is different due to the size and the number of layers in each graphene sample.

For the monolayer graphene sheet deposited at high vacuum under only an applied voltage, we did not observe any rollup of the sheet (see Fig. 2(a)). Figure 4 shows the HRTEM images of bilayer and multi-layer graphene sheets deposited on copper TEM grids in hydrogen atmosphere under an electrostatic field. They clearly indicate that the sheets are scrolled up. In particular, the HRTEM image in Fig. 4(a) shows clearly the edges of two rolled graphene layers forming a tubular structure, while in Fig. 4(b) the HRTEM image shows approximately 16 graphene layers rolled into a tubular structure with a hollow core at the center.

These experimental results can be understood as follows. The hydrogen-exposed monolayer graphene sheet is initially dislodged from the substrate due to the enhancement of the lifting force by the electrostatic field. Subsequently the sheet is scrolled up driven by the stress induced by the adsorption of hydrogen atoms. Since the area of the hydrogen-exposed monolayer graphene sheet is too large for the formation of a SWCNT, the bending of the sheet due to the stress results in the scrolling of the sheet.

Next, we studied the electrostatic deposition of graphene under He and $N_2$ atmospheres for graphene sheets of comparable dimensions. In Fig 5(a) and 5(b), we show the AFM images corresponding to He-exposed monolayer graphene sheet under an electrostatic field for two different samples. There is no indication of the scrolling of the graphene sheet. This is because He atoms do not bond chemically to the carbon atoms. On the other hand, the AFM images shown in Figs. 5 (c) and 5(d) for the electrostatic deposition of graphene under $N_2$ atmosphere for two different samples reveal partial scrolling.

In summary, the following conclusions can be drawn from the results of our experimental studies. The rollup of a monolayer graphene sheet depends on two factors: the dislodging of the sheet from the substrate and the subsequent rollup due to the bending of the sheet induced by the mechanical stress on the sheet. We find that the adsorption of appropriate atomic species on a monolayer graphene sheet alone is not sufficient to induce the roll up of the sheet of large sizes. On the other hand, the combination of the adsorption of appropriate atoms such as hydrogen on a monolayer graphene sheet and the application of an electrostatic field apparently satisfies both requirements and leads to the roll up of the sheet. Our experimental results indicate that the application of an electrostatic field is capable of lifting a monolayer graphene sheet of large dimension from the substrate. This result is expected to eliminate the concern over the size limitation of SWCNTs that could be synthesized using a single GNR as pointed out in Ref. 20, thus circumventing the difficult and infeasible approach of synthesizing SWCNTs of larger diameters using two GNRs.

While the present study is focused on the rollup of a monolayer graphene sheet or graphene layers, our ultimate goal is to demonstrate experimentally the feasibility of synthesizing SWCNTs of pre-defined diameters and chiralities using GNRs. Therefore, we are currently investigating heavy molecular cluster bombardments of graphene to produce GNRs of uniform widths. Preliminary results of producing GNRs of uniform widths using Ga are given in the supplementary information.[29]

ACKNOWLEDGMENT

This work was supported by the U.S. Army, Strategic Missile Defense Command.

FIGURES

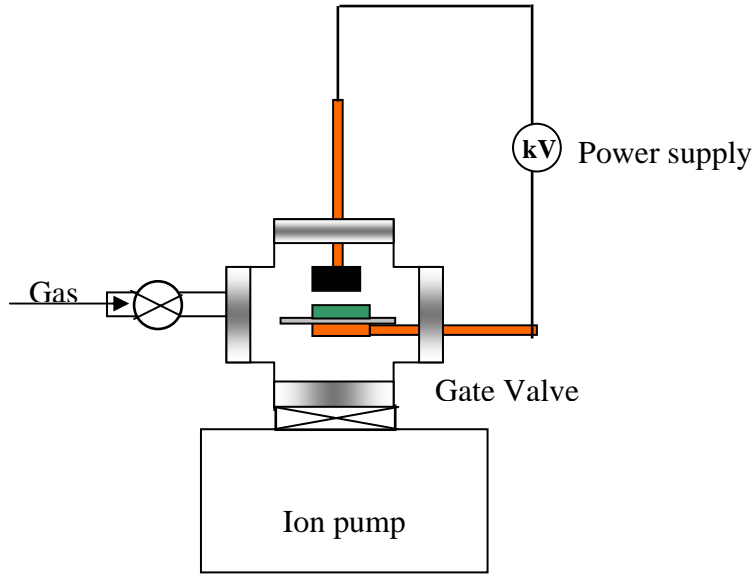

**Figure 1 Setup used for the electrostatic deposition of graphene.** A schematic diagram of the experimental setup used for the electrostatic deposition of graphene sheets under various environmental conditions. In this setup, one side of a HOPG sample is glued to a copper electrode with silver epoxy. The other side of the HOPG sample is cleaved using the scotch tape technique to obtain a clean surface. The second electrode is a 3-mm thick copper plate with a 0.1 mm thick mica sheet placed on it. The substrate of interest is placed over the mica sheet. The electrodes were connected to the terminals of a high voltage source (0-30 kV and 0-10 mA, ER series, Glassman High Voltage, Inc.) via high voltage vacuum feedthroughs.

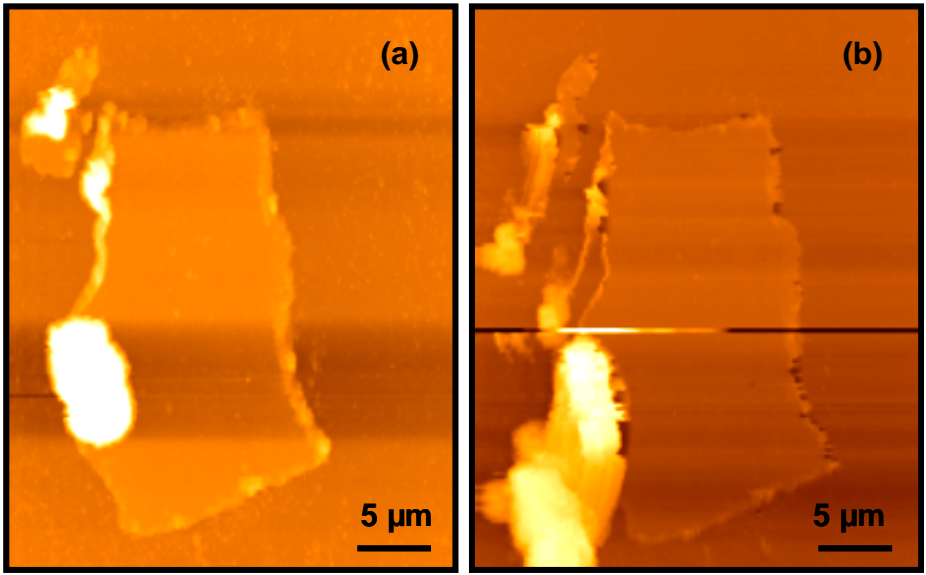

**Figure 2** AFM images of graphene. a, A graphene sheet deposited electrostatically under high vacuum ~$10^{-7}$ torr on a Si/SiO$_2$ substrate. b, The same graphene sheet after exposure to hydrogen at 50 Torr. Both images are very flat, with no indication of any scrolling due to the exposure to hydrogen.

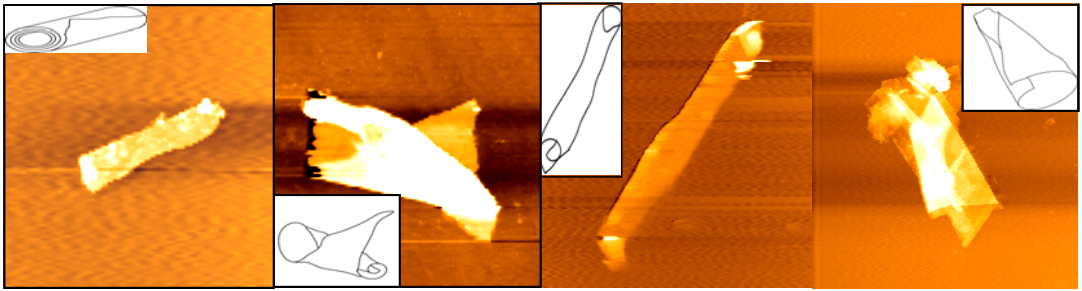

**Figure 3 Electrostatic deposition of graphene under hydrogen atmosphere.** The AFM image of a scrolled graphene sheet deposited under hydrogen atmosphere (50 Torr) using an applied voltage of 450 V. The inset in each figure shows the schematic of the scrolled graphene layer for clarity. The degree of scrolling in each case is different due to the size and the number of layers in each graphene sample.

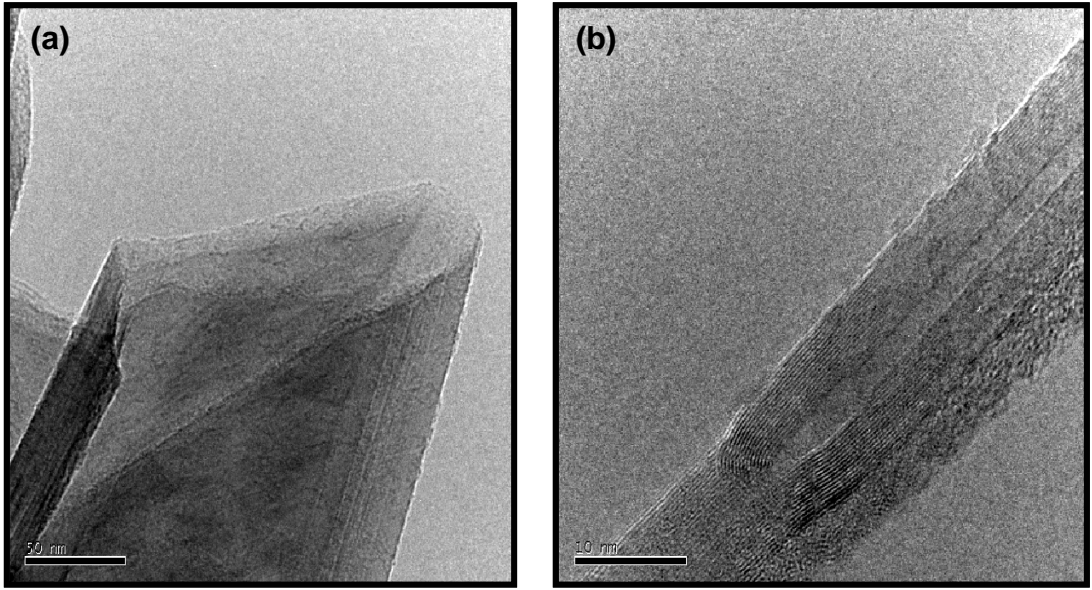

**Figure 4 HRTEM images of rolled graphenes.** The HRTEM images of the rolled graphene sheets deposited on TEM grids under hydrogen atmosphere (50 Torr) using an applied voltage of 450 V; a, rolled bilayer. This clearly shows the edges of two rolled graphene layers forming a tubular structure; b, rolled multilayer showing ~16 graphene layers rolled into a tubular structure with a hollow core at the center.

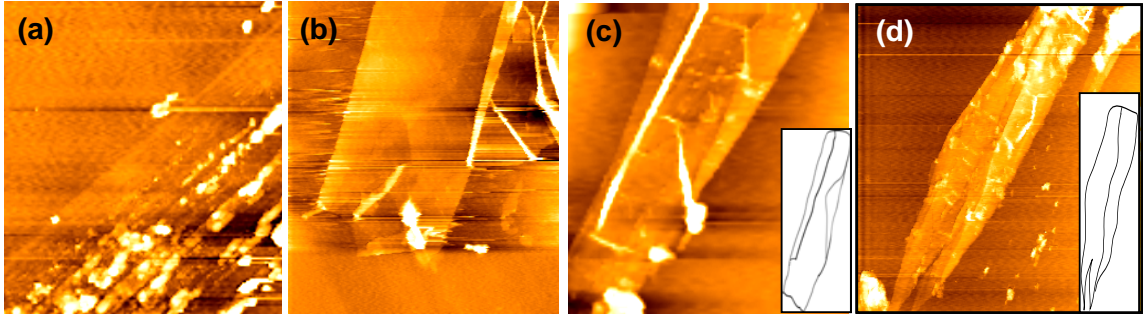

**Figure 5 Electrostatic deposition of graphene under helium and nitrogen atmospheres.** **a** and **b,** The AFM images of graphene sheets deposited under Helium atmosphere (50 Torr) and in the presence of an electrostatic field. Both images indicate that the graphene layers are very flat with sharp edges with no appreciable rolling/scrolling. **c** and **d**, The AFM images of graphene sheets deposited under nitrogen atmosphere (50 Torr) and in the presence of an electrostatic field . They reveal partial scrolling of the graphene.

FIGURE CAPTIONS

**Figure 1 Setup used for the electrostatic deposition of graphene.**

A schematic diagram of the experimental setup used for the electrostatic deposition of graphene sheets under various environmental conditions. In this setup, one side of a HOPG sample is glued to a copper electrode with silver epoxy. The other side of the HOPG sample is cleaved using the scotch tape technique to obtain a clean surface. The second electrode is a 3-mm thick copper plate with a 0.1 mm thick mica sheet placed on it. The substrate of interest is placed over the mica sheet. The electrodes were connected to the terminals of a high voltage source (0-30 kV and 0-10 mA, ER series, Glassman High Voltage, Inc.) via high voltage vacuum feedthroughs.

**Figure 2 AFM images of graphene.**

a, A graphene sheet deposited electrostatically under high vacuum ~$10^{-7}$ torr on a $Si/SiO_2$ substrate. b, The same graphene sheet after exposure to hydrogen at 50 Torr. Both images are very flat, with no indication of any scrolling due to the exposure to hydrogen.

**Figure 3 Electrostatic deposition of graphene under hydrogen atmosphere.**

The AFM image of a scrolled graphene sheet deposited under hydrogen atmosphere (50 Torr) using an applied voltage of 450 V. The inset in each figure shows the schematic of the scrolled graphene layer for clarity. The degree of scrolling in each case is different due to the size and the number of layers in each graphene sample.

**Figure 4 HRTEM images of rolled graphenes.**

The HRTEM images of the rolled graphene sheets deposited on TEM grids under hydrogen atmosphere (50 Torr) using an applied voltage of 450 V; a, rolled bilayer. This clearly shows the edges of two rolled graphene layers forming a tubular structure; b, rolled multilayer showing ~16 graphene layers rolled into a tubular structure with a hollow core at the center.

**Figure 5 Electrostatic deposition of graphene under helium and nitrogen atmospheres.**

**a** and **b**, The AFM images of graphene sheets deposited under Helium atmosphere (50 Torr) and in the presence of an electrostatic field. Both images indicate that the graphene layers are very flat with sharp edges with no appreciable rolling/scrolling. **c** and **d**, The AFM images of graphene sheets deposited under nitrogen atmosphere (50 Torr) and in the presence of an electrostatic field . They reveal partial scrolling of the graphene.